\documentclass[11pt, a4paper]{article}
\usepackage{jheppub}
\usepackage{graphicx}
\usepackage{epsfig,amsmath,amsthm}
\usepackage{epstopdf}

\newcommand{\bea}{\begin{eqnarray}}
\newcommand{\eea}{\end{eqnarray}}
\newcommand{\be}{\begin{equation}}
\newcommand{\ee}{\end{equation}}
\newcommand{\ben}{\begin{equation*}}
\newcommand{\een}{\end{equation*}}
\newcommand{\bealn}{\begin{align*}}
\newcommand{\eealn}{\end{align*}}
\newcommand{\beal}{\begin{align}}
\newcommand{\eeal}{\end{align}}

\newcommand{\non}{\nonumber}
\renewcommand{\vec}[1]{\mathbf{#1}} 

\def\bi{\begin{itemize}} \def\ei{\end{itemize}}

\makeatletter
\let\over=\@@over \let\overwithdelims=\@@overwithdelims
\let\atop=\@@atop \let\atopwithdelims=\@@atopwithdelims
\let\above=\@@above \let\abovewithdelims=\@@abovewithdelims
\makeatother

\title{Leading anomalies, the drift Hamiltonian and the relativistic two-body system}

\author{Bernard M. Nabet and Barak Kol\\
{\it Racah Institute of Physics, Hebrew University, Jerusalem 91904, Israel} \\
}

\abstract{We suggest to solve for the motion of the two body problem in General Relativity by identifying the leading violation of conserved quantities, referred to as (relativistic) anomalies, ordered by the post-Newtonian order at which they appear. This differs from the standard procedure of obtaining the full solution up to a prescribed order. We find that the reduced Hamiltonian which describes the drift in the space of conserved quantities is given by the average of the perturbation Hamiltonian. Using this approach the averaging is done prior to the derivation of time evolution, thereby economizing the computation. The computations become similar to those in the Hamilton-Jacobi method, while staying in the more comfortable setting of the Hamiltonian formulation. We apply this approach of leading anomalies and the drift Hamiltonian to the binary problem and treat several perturbations: 1PN, spin-orbit and spin-spin. On the way we discuss the interpretation of the Laplace-Runge-Lenz vector as a generator of scale-preserving conformal transformations in momentum space.}

\begin{document}
\maketitle
\section{Introduction}


Solving the two-body problem in Einstein's gravity is important both intrinsically and for the ongoing worldwide effort to observe gravitational waves.  First, solutions are helpful for gravitational wave detection since they are incorporated in the design of the signal detection filters, and secondly, once gravitational waves are detected the solutions would be essential for signal interpretation.

The post-Newtonian domain enables analytical study of the two body problem, and was widely studied from the very first days of General Relativity, see the review \cite{BlanchetRev13}. This field of research can be broken down to two parts. The first is to obtain the two body effective action through the elimination of the (Einstein) gravitational field, and the second is to solve the resulting equations of motion. The first part probably attracted most of the research attention so far, and is not the subject of this work. Its objective is to obtain relativistic corrections to the Newtonian two-body action. The pioneering works were \cite{LorentzDroste17,EIH38}\footnote{
Clearly, while \cite{LorentzDroste17} was earlier and in several respects closer to the modern methods it was presumably unknown to the authors of \cite{EIH38}.}
 and some of the key concepts in the consequent evolution of the field were the action formulation \cite{LorentzDroste17} or alternatively the Hamiltonian one \cite{SchaeferRev11}; diagrammatic tools starting with \cite{OOKH73,DamourFarese95} and culminating in the effective field theory approach and standard Feynman diagrams of \cite{GoldbergerRothstein04};  renormalization and regularization, see e.g. \cite{Bel:1981be2,GoldbergerRothstein04}; a non-relativistic decomposition of the Einstein gravitational field \cite{Kol2007} with roots in the early days of GR; and finally the choice of gauge for the gravitational field, where popular choices include the harmonic gauge, e.g. \cite{BlanchetRev13}, and the ADM gauge, e.g. \cite{SchaeferRev11}.

This paper is concerned with solving the effective equations of motion for the two bodies, which are the second part in the post-Newtonian analysis. Usually the equations of motion are solved up to a prescribed post-Newtonian order of accuracy.  Here instead we consider the quantities conserved by the Newtonian problem and seek their leading violation (and possibly corrections) at whichever order it may appear.   We refer to the PN violation of conserved quantities as \emph{relativistic anomalies}, or anomalies in short. The standard usage of the term anomaly is within Quantum Field Theory, where it describes the violation of a classically conserved quantity due to quantum effects. Here we generalize the term to apply to any quantity which is conserved in the a theory, but broken in the more general one, and so here relativistic effects replace the quantum effects in the ordinary context. However, whereas the coefficient of quantum anomalies are typically topological and hence integral, this is not the case for relativistic anomalies.\footnote{Our usage of the term anomaly is unrelated to its other use in celestial mechanics as a parameter that defines the position along an orbit, namely the mean anomaly, the eccentric anomaly and the true anomaly.} The motivation of the anomaly approach is that a high order, yet leading, anomaly is in some sense a more dramatic and physical effect than a high order correction of the same order for some other anomaly.

The anomaly approach leads us to consider the dynamics in the reduced space of conserved quantities where relativistic corrections induce a slow, average drift of the conserved quantity. As we show in section \ref{sec:method} an appropriate reduced \emph{drift Hamiltonian} can be defined for this motion and it is nothing but the time-average of the original perturbation Hamiltonian. Mathematically the averaging integrals are conveniently performed by an analytic continuation and a residue method such as in appendix \ref{app:Residu}. 

 In the following sections we proceed to apply these concepts to the post-Newtonian two body problem. We start in part \ref{sec:newton} by reviewing the Newtonian two-body problem with an emphasis on conserved quantities and a special discussion of the Laplace-Runge-Lenz (LRL) vector as a generator of scale-preserving conformal transformations in momentum space\footnote{
 A subgroup of conformal transformation which preserves a momentum scale associated with the energy, and hence the symmetry does not include scale invariance.}
 in section \ref{sec:LRL}. 
 Next in part \ref{part:anomalies} we treat the following relativistic perturbations: 1PN in section \ref{sec:H1PNr},  spin-orbit in section \ref{sec:Hso} and finally spin-spin in section \ref{sec:Hss}. In each case we present the perturbation Hamiltonian, compute from it the drift Hamiltonian and then use it to compute the appropriate leading  anomalies.

\subsection{Summary and discussion}

Our main results are
\bi
\item The proposed anomaly approach -- explained in the introduction.
\item The definition of the drift Hamiltonian (\ref{H-drift}) and its usage in the post-Newtonian context.
\item A novel economic computation of the the post-Newtonian periapsis shift viewed as the 1PN anomaly in the LRL vector (\ref{A1pn}).
\item A full treatment of the spin-orbit and spin-spin interactions within our approach including the computation of several anomalies (\ref{Ldot}, \ref{S1dot}, \ref{SO-Adot}, \ref{Ldotss}, \ref{S1dotss}, \ref{SS-Adot}).  
\ei
In addition we present a rather detailed discussion of the the symmetry underlying the LRL vector \cite{Pauli26,Fock1935,Bander1966, CaronHuotAmplitudes14} namely conformal symmetry in momentum space in section \ref{sec:LRL}. 

Our results are confirmed to be correct as they agree with known expressions. The drift Hamiltonian turns out to appear already in the literature \cite{Moser1970,Reeb52}, where it was called the averaging method, yet this appears to be its first application to the post-Newtonian context. While the above mentioned expressions for the relativistic anomalies are known, the current derivations are novel.  It is interesting to compare the current method with others. On the one hand it is more economical than those which first compute the time variation and only then average. On the other hand when compared with  \cite{HergtShahSchaefer13} which is a rather elegant treatment based on the Hamilton-Jacobi formalism we find that in cases where both methods apply the computational task is comparable, and in particular both require to perform averaging integrals. Yet our method avoids the Hamilton-Jacobi formalism and remains in the more intuitive and familiar Hamiltonian formalism.

It would be interesting to apply our method to additional anomalies. In particular it remains to incorporate the dissipative effects of radiation reaction which break the conservation of energy, angular momentum (at 2.5PN) as well as center of mass momentum. Doing that at a Hamiltonian level should require field doubling \cite{Galley09,BirnholtzHadarKol13}.

\section{The averaging method}
\label{sec:method}

Quite generally for a perturbed mechanical system,
classical works first compute the time derivative $\dot{X}$, where $X$ is some dynamics variable, and then compute the secular variation by averaging over a complete Newtonian orbit. This can be done either at the level of the equations of motion, or in a Hamiltonian formulation. In the latter we have \be
 H= H_0 + H' ~,
 \ee
 where $H_0$ is the unperturbed Hamiltonian and $H'$ is the perturbation. Now the perturbed time variation of any dynamic variable $X$ is given by \be
 \dot{X} - \dot{X}_0 = \{ X,H' \} ~,
 \label{x-dot}
 \ee
where $\{X,H\}$ is the Poisson bracket and $\dot{X}_0 := \{X,H_0\}$ is the unperturbed time variation.

We wish to consider quantities which are conserved in the unperturbed system, denoted here by $A_i$. If $A$ is one such quantity then by definition $\{A,H_0\}=0$ and so  (\ref{x-dot}) becomes \be
\dot{A} =  \{ X,H' \} ~.
\label{A-dot}
\ee
Given a specific unperturbed orbit $X_0=X_0(t;A_i)$ and substituting into $\dot{A}$ one obtains $\dot{A}=\dot{A}(t;A_i)$, which in turn decomposes into oscillatory terms and a constant term. The latter is usually of greater interest as it describes the long-term drift in $A$, also known as the secular variation. This constant term can be extracted through a time average $\langle \dot{A} \rangle$ over a full period, denoted by the angled brackets. The set of all drifts $\langle \dot{A} \rangle$ defines a dynamical system on the space of conserved quantities. One may wonder whether this dynamical system is Hamiltonian, namely, whether it can be defined through a reduced Hamiltonian function $H_d=H_d(A_i)$. In fact, by interchanging the averaging operation with the Poisson bracket we have \be
\langle \dot{A} \rangle = \langle \{ A,H' \} \rangle =  \{ A,\langle H' \rangle \}   ~ \label{Adotmethod},
\ee
from which recognize that $H_d$ exists and is given by \be
H_d := \langle H' \rangle ~,
 \label{H-drift}
\ee
that is, the drift Hamiltonian is nothing but the perturbation Hamiltonian time-averaged over the orbits of $H_0$. We note that while $H'=H'(X_J)$ is a function of the original dynamical variables $X_J$, $H_d$ is constant over orbits by construction, and hence it depends only on the conserved variables, namely $H_d=H_d(A_i)$.

After obtaining this general result we found that it was already described in \cite{Moser1970} and termed the averaging method. There it was applied in combination with topological methods to show the existence of periodic solutions in the restricted three-body problem. \cite{Moser1970} mentioned that the method had already been found in \cite{Reeb52}, yet  the Kepler problem was not treated there. We are unaware of an application of this method in the post-Newtonian context prior to the current work.

\part{Newtonian symmetries}
\label{sec:newton}

\section{The Kepler problem}

\begin{figure}[!hb]
\includegraphics[width=6cm]{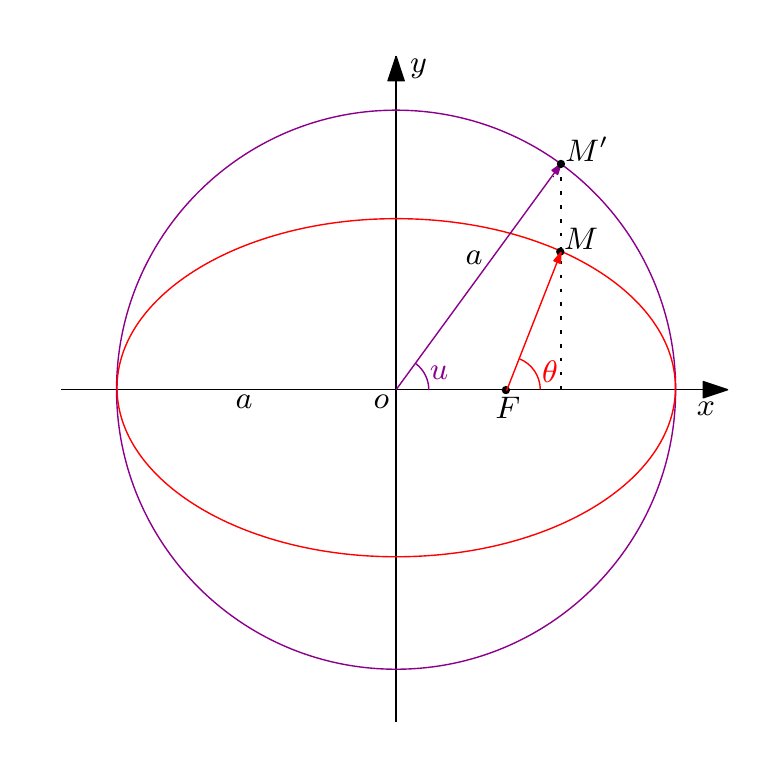}\centering
\caption{The eccentric anomaly of point $M$ is the angle $u$. M' is determined by the intersection between a circle of radius $a$ and the line vertical to the major axis ($x-axis$) and passing through point M.}\label{fig:eccentricanomaly}
\end{figure}

We are considering a two-body system composed of two masses $m_1$ and $m_2$ under gravitational interaction. The Lagrangian of the two body system is given by
\be
L_{0}= \frac{1}{2} m_1 \vec{v_1}^2+\frac{1}{2} m_2 \vec{v_2}^2 + G\frac{m_1 m_2}{r}
\ee
in the center-of-mass frame, the relative Lagrangian is
\be
L_{0} = \frac{1}{2} \mu \vec{v}^2+G\frac{m_1 m_2}{r}~,
\ee
where $\mu=\frac{m_1 m_2}{m_1 + m_2}$ is the reduced mass and $\vec{v}=\vec{v_2}-\vec{v_1}$ and the relative Hamiltonian is
\be
H_{0}= \frac{\vec{p}^2}{2 \mu}  -G\frac{m_1 m_2}{r}~.
\ee
The Lagrangian is invariant under time translation an rotations. Noether's theorem implies the conservation of energy and angular momentum conservation which are given by
\begin{align*}
E &= \frac{1}{2} \mu \vec{v}^2-G\frac{m_1 m_2}{r}~,\\
\vec{L} &= \mu \vec{r}\times \vec{v}~.
\end{align*}
 L conservation means that the motion is planar and using polar coordinate we have
\begin{align}
\vec{v}^2 &= \dot{r}^2+r^2 \dot{\theta}^2~,\\
\vec{L}&=\mu r^2 \dot{\theta} \vec{\hat{e}_Z}~,\\
E &= \frac{1}{2} \mu\dot{r}^2 + \frac{L^2}{2 \mu r^2} - \frac{Gm_1m_2}{r}~.  \label {E}
\end{align}
Now the radial equation becomes
\be
\dot{r}^2 = \frac{2}{\mu} \left(E - \left[\frac{L^2}{2\mu r^2}- \frac{Gm_1m_2}{r}\right]\right) \label{rdot1}~,
\ee
From now, we will use reduced physical quantities (the reduced energy is defined by $\tilde{E}:=\frac{E}{\mu}$ and the reduced angular momentum by $\tilde{\vec{L}}=\frac{\vec{L}}{\mu}$ reduced angular momentum) and denote them by E, $\vec{L}$ omitting the tilde symbol.(\ref{rdot1}) becomes
\be
\dot{r}^2 = 2\left(E - \left[\frac{L^2}{2 r^2}- \frac{\alpha}{r}\right]\right) \label {rdot} .
\ee
with $\alpha=G(m_1 +m_2)$. The solution is
\begin{align}
\theta(r) &= \int \frac{\frac{L}{r^2}}{\sqrt{2( E - [\frac{L^2}{2r^2}- \frac{\alpha}{r}])}}dr~,\\
t(r) &= \int_{r_0}^{r} \frac{dr}{\sqrt{2( E - [\frac{L^2}{2r^2}- \frac{\alpha}{r}])}}~.
\end{align}
For bounded motion ($E < 0$), the solutions are elliptical closed orbits with eccentricity parameter $e$
\be
r(\theta) = \dfrac{L^2}{\alpha(1+e\cos\theta)} \qquad \text{with}~~e=\sqrt{1+\frac{2EL^2}{\alpha^2}} \leq 1~.\\
\ee
\\
Using $a$ the semi-major axis of the ellipse, we finally have
\begin{align}
r(\theta) &= \dfrac{L^2}{\alpha(1+e\cos\theta)} = \dfrac{ a(1-e^2) }{(1+e\cos\theta)} \label{rtheta}~,\\
E &= -\dfrac{\alpha}{2a}~,\\
L^2 &= a\alpha(1-e^2) = \frac{\alpha^2}{(-2E)} (1-e^2) \label{L2}~,\\
\omega_0 &= \dfrac{2\pi}{T}=\left(\dfrac{\alpha}{a^3}\right)^{\frac{1}{2}}~. \label{omega0}
\end{align}
It is sometimes more convenient to reparameterize (\ref{rtheta}) and replace $\theta$ by another angle $u=u(t)$ known as ``the eccentric anomaly" shown in figure \ref{fig:eccentricanomaly} such that
\be
r=a(1-e\cos u)~, \label{u1}
\ee
\\
We then get the following equations
\begin{align}
x &=a(\cos u-e)~, \\
y &=a\sqrt{1-e^2}\sin u = b \sin u~,\\
\omega_0t &=u -e \sin u~. \label{u2}
\end{align}
with $b$ the semi-minor axis of the ellipse.

\subsection{LRL Vector}
There is an additional conserved quantity $\vec{A}$, the Laplace-Runge-Lenz (LRL) vector, see e.g.\cite{Goldstein0}, defined as (using reduced $\vec{A}, \vec{p}$ and $\vec{L}$)
\begin{equation}
\vec{A}=\vec{p} \times \vec{L} -  \alpha \frac{\vec{r}}{r}~.
\label{def:LRL}
\end{equation}
It is named after Pierre-Simon Laplace who defined it in 1799 with a clear and reasonably complete physical reasoning , Wilhelm Lenz who used it in 1924 to compute the Hydrogen atom spectrum in the old quantum theory and finally Carl Runge to whose 1919 book Lenz referred \cite{GoldsteinLRL1}. In fact, it was already known to Jakob Hermann and Johann I. Bernoulli in 1710 \cite{GoldsteinLRL2}. See \cite{MacKaySalour} for recent work on the subject and a useful review of it.
This vector is always in the orbital plane, indeed from the definition of $\vec{A}$ we have
\be
\vec{A} \cdot \vec{L} = 0 \label{A1}~.
\ee
since $\vec{L}$ is perpendicular to both $\vec{p}\times \vec{L}$ and $\vec{r}$. It is pointing in the direction of the periapsis (semi major axis of the ellipse) and its magnitude is given by $A=\alpha e$ or using (\ref{L2})
\be
 A^2=\alpha^2+2EL^2 \label{A2}~.
\ee

\subsection{Hodographs}
\begin{figure}[!hb]
\includegraphics[width=10cm]{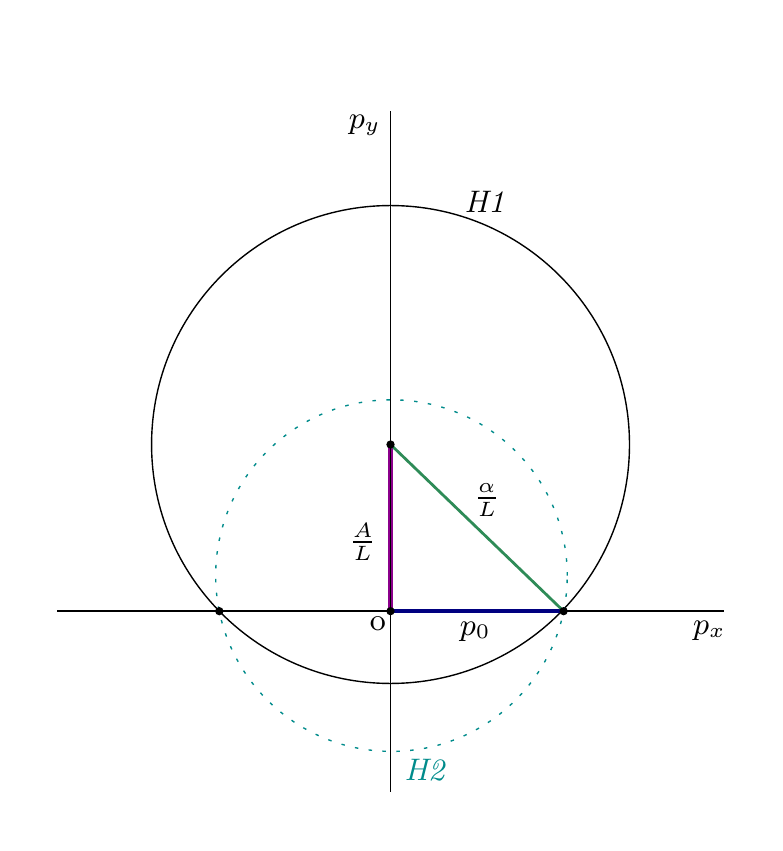}\centering
\caption{H1 is the hodograph of a bounded orbit of energy $p_0=\sqrt{-2E}$ with given values for $\vec{L}$ and $\vec{A}$ which determine eccentricity. The hodograph H2 has same energy but different eccentricity.}\label{fig:hodographdessin}
\end{figure}

Hamilton showed that in momentum space the trajectories of the two-body problem are perfect circles (such trajectories are called hodographs).We can show it using the LRL vector. We start with the identity
\be
\vec{L} \times \vec{A}=\vec{L}\times(\vec{p}\times\vec{L})-\alpha\frac{\vec{L}\times\vec{r}}{r}=\vec{p}L^2-\alpha\frac{\vec{L}\times\vec{r}}{r}~. \label{LxA}
\ee
Without loss of generality, we can choose $\vec{L}$ along the z-axis that is $\vec{L}=(0,0,L)$ and the semi-major axis along the x-axis that is $\vec{A}=(A,0,0)$ and since the motion lies in the $x y$ plane $\vec{r}=(x,y,0)$ and $\vec{p}=(p_x,p_y,0)$, (\ref{LxA}) gives
\begin{align}
p_xL^2 &=\alpha\frac{yL}{r}~,\\
p_yL^2 &=AL-\alpha\frac{xL}{r}~,
\end{align}
and then
\be
{p_x}^{2}+\left(p_y-\frac{A}{L}\right)^2=\left(\frac{\alpha}{L}\right)^2~. \label{hodograph_circle}
\ee
Thus in the momentum space, as shown in figure \ref{fig:hodographdessin}, bounded orbits are represented by circles of radius $\alpha/L$ centered on $(0, A/L)$.
The eccentricity $e$ of the orbit is given by the ratio between the position of the center and the radius of the circle ($e=\frac{A}{\alpha}$) and the intersection point ($p_0$,0) of any circle orbit with the $p_x$-axis gives the orbit energy since from (\ref{A2})
\be
{p_0}^{2}=\left(\frac{\alpha}{L}\right)^2-\left(\frac{A}{L}\right)^2=-2E \label{hodograph_p0}~.
\ee

We see that there is a family of hodographs with the same energy but with varying eccentricity. In the next section we shall see that this transformation is in fact a symmetry of the mechanical system.

\subsection{Conserved quantities}
We have identified 3 constants of motion: two vector constants $\vec{L}$ and $\vec{A}$ and a scalar E. The seven quantities composing those constants are not independent since we have two relations between $\vec{L}$, $\vec{A}$ and E : the magnitude of $\vec{A}$ can be determined from L and E using (\ref{A2}) and since $\vec{A}$ must be perpendicular to $\vec{L}$ (\ref{A1}). There is thus only five independent constants of the motion.
 
The conservation of $\vec{A}$ implies that our problem has another symmetry in addition to time translation (conservation of E) and space rotation (conservation of $\vec{L}$). Since this symmetry is not evident in the Lagrangian $L_0$ it has been called ``hidden symmetry". For spinning objects, the spin $\vec{S_1}$, $\vec{S_2}$ are conserved too in the Newtonian limit.

\section{LRL vector and higher symmetry}
\label{sec:LRL}

The existence of an unforeseen conserved quantity, the LRL vector (\ref{def:LRL}), is a sign of enhanced symmetry. In this section we will study that symmetry.

The Lie algebra of symmetry generators is given by the Poisson brackets of the corresponding conserved quantities. For bound orbits ($E < 0$) we have 
\begin{align}
\{L_i,L_j\} &= \epsilon_{ijk}L_k~,\\
\{A_i,L_j\} &= \epsilon_{ijk}A_k~,\\
\{A_i,A_j\} &= -p_0^2\, \epsilon_{ijk}L_k ~.
\label{LA-Poisson}
\end{align}
where $p_0:=\sqrt{-2 E}$ and recall that $E,\, \vec{L}$ and $\vec{A}$ are reduced quantities. $p_0$ can be eliminated from this Poisson algebra by rescaling $\vec{A} \to p_0\, \vec{A}$. Then we recognize it to be the $SO(4)$ algebra, namely the generators of rotations in 4 dimensions. This symmetry includes the manifest $SO(3)$ algebra of 3d rotations generated by $L_i$.\\

The enhanced symmetry raises the following two related questions \bi
 \item Why does it exist?\\
 	Namely, from what point of view could one \emph{anticipate} it?

 \item Physical or geometrical interpretation.\\
  	$SO(4)$ is the group of 4d rotations. Could one recast the problem in convenient variables such that the symmetry is simplified to ordinary rotations in that space?    
 \ei

Recently interesting answers to these questions were suggested by Caron-Huot and Henn \cite{CaronHuotAmplitudes14}, according to which the enhanced symmetry is \emph{a subgroup of the conformal transformations in momentum space}. This suggestion is motivated by some similarity with dual conformal invariance which appears in modern studies of integrability in ${\cal N}=4$ supersymmetric 4d field theory. The line of argument can be described as follows. One considers the ladder diagrams in a model considered by  Wick and Cutkosky \cite{WickCutkosky54} for electron-proton scattering, which is the same system whose bound state is the Hydrogen atom with its Kepler dynamics. The 1-loop ladder diagram is recognized to enjoy a conformal symmetry in momentum space. This is made explicit by Dirac's conformal compactification of Minkowski space in 6d \cite{Dirac1935}. Finally, the  conformal transformations are limited to the subgroup of the  conformal group $SO(4) \subset SO(4,2)$ which preserves the two incoming momenta.

These intuitions are imported from Quantum Field Theory to a non-quantum non-relativistic problem in mechanics (no fields). Clearly there should be an intrinsic way to think about this symmetry, namely one which remains within in the original context of the problem. In this section we will attempt to study the problem step by step and gain some intuition into this interpretation of the symmetry.

\subsection{Phase space symmetry}

We wish to determine the generators of the symmetry in phase space. The computation of the infinitesimal variation of $\vec{r}$ and $\vec{p}$ corresponding to $\vec{A}$ conservation can be done using Poisson brackets
\begin{align}
\delta_j r^ i &= \{A^j , r^i \}= \{ r^j p^2 -p^j (p^l\cdot r^l)-\alpha \frac{r^j}{r},r^i \} \label{deltajri}~,\\
\delta_j p^ i &=\{A^j , p^i \}= \{ r^j p^2 -p^j (p^l\cdot r^l)-\alpha \frac{r^j}{r},p^i \} \label{deltajpi}~.
\end{align}
where we used $\vec{p} \times \vec{L} = \vec{r} p^2 -\vec{p}(\vec{p}\cdot\vec{r})$.

After computation, see Appendix \ref{app:LRLtransformation}, we finally find the infinitesimal transformation corresponding to LRL vector conservation \cite{LevyLeblond}
\begin{align} 
\delta_j r^ i &=  \{A^j , r^i \}=[-2r^jp^i+r^i p^j +(\vec{r}\cdot\vec{p})\delta_{ij}]~,\\
\delta_j p^ i &= \{A^j , p^i \}=[p^2 \delta_{ij}-p^j p^i  
-\alpha (\frac {\delta_{ij}}{r}-\frac{r^j r^i}{r^3})] ~. \label{DeltaLRL}
\end{align}
The physical interpretation of these transformations is not manifest. To get some intuition and following a computation from \cite{Mostowski2010} we shall show that, for a system with orbit motion in the $(x,y)$ plane and angular momentum $L=L\hat{\vec{z}}$ along z-axis, the infinitesimal transformation generated by $A_y$ is equivalent to a specific variation of the eccentricity. Working in the momentum space, we have (\ref{DeltaLRL})
\begin{align}
\delta_y p_x &=(-p_x p_y+\alpha\frac{x y}{r^3})~,\\
\delta_y p_y &= (p^2-{p_y}^2-\alpha(\frac{1}{r}-\frac{y^2}{r^3}))=({p_x}^2-\alpha\frac{x^2}{r^3})~.
\end{align}
Using (\ref{u1}, \ref{u2}) we compute
\begin{align}
\frac{du}{dt}&=\frac{\omega_0}{1-e \cos u}~,\\
p_x &=\frac{dx}{dt}=\frac{dx}{du}\frac{du}{dt}=-\frac{a\omega_0 \sin u}{1-e \cos u} \label{pxellipse}~,\\
p_y &=\frac{dy}{dt}=\frac{dy}{du}\frac{du}{dt}=\frac{b\omega_0 \cos u}{1-e \cos u}~.
\end{align}
and substituting back, using \ref{omega0} again we find
\begin{align}
\delta_y p_x &= ab{\omega_0}^2\left(\frac{\cos u\sin u}{(1-e \cos u)^2}+\frac{\sin u(\cos u-e)}{(1-e \cos u)^3}\right) \label{deltapx}~,\\
\delta_y p_y &= a^2{\omega_0}^2 \left(\frac{\sin^2 u}{(1-e \cos u)^2}-\frac{(\cos u-e)^2}{(1-e \cos u)^3}\right) \label{deltapy}~.
\end{align}

Let us show that these infinitesimal transformation correspond to a variation of the magnitude of the eccentricity (at constant $\omega_0$ and $t$), that is $\delta p_i =\frac{\partial p_i}{\partial e}\delta e$.\\
From (\ref{u2})
\be
\frac{\partial u}{\partial e}=\frac{\sin u}{1-e\cos u}~,
\ee
Varying (\ref{pxellipse}) we find
\begin{align}
\delta p_x &=\delta e~a\omega_0\left(-\frac{\cos u \sin u}{(1-e \cos u)^2}+\frac{\sin u \cos u (e\cos u-1)+e\sin^2 u}{(1-e \cos u)^3}\right)\\
&=-\delta e~a\omega_0\left(\frac{\cos u \sin u}{(1-e \cos u)^2}+\frac{\sin u(\cos u-e)}{(1-e \cos u)^3}\right)~,
\end{align}
and (after a more tedious calculation)
\begin{align}
\delta p_y &=-\delta e\frac{a\omega_0}{\sqrt{1-e^2}}\left(\frac{\sin^2 u}{(1-e \cos u)^2}-\frac{(\cos u-e)^2}{(1-e \cos u)^3}\right)~.
\end{align}

Actually, to reproduce (\ref{deltapx}, \ref{deltapy}) exactly the variation should not be taken with respect to the eccentricity $e$ but rather a function of it.
Indeed this can be done by changing variables from $e$ to $\psi$ as follows
\be
e=\sin \psi ~, \qquad \sqrt{1-e^2}=\cos \psi ~. \label{etatransformation}
\ee
Thus the symmetry generated by $A_y$ is equivalent to an infinitesimal variation of the periodic variable $\psi/p_0=\frac{1}{p_0}\arcsin e$, where $p_0 \equiv a \omega_0$, namely it is an infinitesimal rotation. The other components of $\vec{A}$ have an analogous interpretation.

\subsection{The hypersphere}

The $SO(4)$ symmetry algebra (\ref{LA-Poisson}) leads us to expect the existence of a change of dynamical variables where the symmetry  simplifies to rotations in 4d. This expectation is partially realized by the demonstration in the last subsection that the components of $\vec{A}$ generate rotations in variables such as $\psi$ defined in (\ref{etatransformation})

\begin{figure}[!hb]
\includegraphics[width=8.5cm]{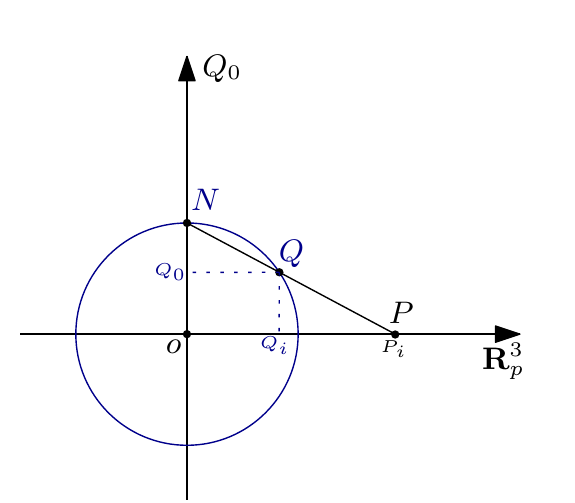}\centering
\caption{The point P in the momentum phase is projected to the point Q on the hypersphere.}\label{fig:stereographicdessin}
\end{figure}

A geometrical explanation of the 4d rotation symmetry was given by Fock in 1935 \cite{Fock1935}  for the similar problem of quantum states of the hydrogen atom, see also Bander and Itzykson \cite{Bander1966}.
Starting from the usual 3d momentum space $\mathbb{R}_p^3$ we can construct a 4D momentum space by adding a fourth dimension ($Q_0$-axis) and projecting $\mathbb{R}_p^3$ on the surface of the 3-sphere $S^3_Q$ centered at the origin and of radius $p_0=\sqrt{-2E}$ using stereographic projection. More specifically given a point $P=(0,p_i)$ of an hodograph of energy $\frac{-p_0^2}{2}$ on the 3D momentum space, the coordinate of its stereographic projection point $Q=(Q_0,Q_i)$ on the 3-sphere of radius $p_0$ are given by (see figure  \ref{fig:stereographicdessin})
\be
Q_0=\frac{p^2-p_0^2}{p^2+p_0^2} p_0 ~,\qquad Q_i=\frac{2p_0^2}{p^2+p_0^2}p_i~. \label{stereographic}
\ee
It turns out that the Kepler problem maps to a free particle on the $S^3_Q$ hypersphere. In order to gain some intuition into this fact we will show that hodographs in the momentum space are mapped to sets on $S^3_Q$ which are symmetric with respect to reflection. In fact these are great circles in $S^3_Q$ and thus a transformation from one hodograph to another one with same energy in the momentum space is equivalent to a transformation from one great circle to another great circle on $S^3_Q$ that is a 4D rotation or an SO(4) symmetry.

Now, following an idea from \cite{CMMarle}, let us introduce the Euclidian geometry concept of ``power of a point" $P_C (M)$ which gives the relative distance of a given point from a relative circle.
Given a point M, a circle C of center O and radius r, and A and B the two intersection points of any ray emanating from M with the circle C we have
\be
P_C (M)=\overrightarrow{MA}\cdot\overrightarrow{MB}=(MO)^2-r^2~,
\ee 
In particular, if M is inside the circle, $P_C (M)=- \|\overrightarrow{MA}\|\cdot\|\overrightarrow{MB} \|~\le 0$. Therefore, for any two points A and B of a same hodograph H in the 3D momentum space (\ref{hodograph_circle}, \ref{hodograph_p0}), we have
\be
P_H (O)=\overrightarrow{OA}\cdot\overrightarrow{OB}=\left(\frac{A}{L}\right)^2-\left(\frac{\alpha}{L}\right)=-p_0^2~, \label{power_hodograph}
\ee
Thus the origin O is always inside the hodographs (for bounded orbits $E \le 0$) so if we have $\overrightarrow{OA}>0$ we necessarily have $\overrightarrow{OB}<0$.
We can now show that their stereographic projection points $Q^A$ and $Q^B$ will be reflections, symmetric relative to the origin O.\\
We have
\be
\overrightarrow{OA}\cdot\overrightarrow{OB}=P_H (O)=-p_0^2=-ab \quad \text{with}\quad a=\|\overrightarrow{OA}\|,~b=\|\overrightarrow{OB}\|
\ee
and thus
\begin{align}
Q^A_0&=\frac{a^2-ab}{a^2+ab} \sqrt{ab}=\frac{ab-b^2}{ab+b^2}\sqrt{ab}=-Q^B_0~,\\
\overrightarrow{Q^A} &= \frac{2(ab)}{a^2+ab}~\overrightarrow{OA}= \frac{2(ab)a}{a^2+ab}=\frac{2(ba)b}{ab+b^2}=-\overrightarrow{Q^B}~.
\end{align}
Since the origin O is inside any hodograph, by taking ray emanating from O we can cover the whole hodograph and then map the great circle on $S^3_Q$. This mapping between hodographs and great circle on $S^3_Q$ can be demonstrated more rigorously showing that a free particle hamiltonian on $S^3_Q$ is mapped to a Kepler potential Hamiltonian on the momentum space \cite{Moser1970}. Since any hodograph on momentum phase is mapped to a great circle on $S^3_Q$, a transformation from one hodograph to another one with same energy (constant $p_0$) in the momentum space is equivalent to a transformation from one great circle to another great circle on $S^3_Q$ that is a 4D rotation or SO(4) symmetry! We thus understood in which space the SO(4) symmetry of our problem is taking place.

Now we can understand the exact transformation behind the LRL vector/angular momentum conservation
\begin{itemize}
\item we start from an arbitrary hodograph H of energy  $\frac{-p_0^2}{2}$  in the momentum space
\item we make a stereographic projection to the 3-sphere of center O and radius $p_0$ and obtain a great circle G
\item we make a 4D rotation on the 3-sphere to obtain another great circle G' (without loss of generality, we will only rotate between the new $Q_0$-axis and $1$-axis to simplify computation)
\item we finally project back the new great circle on the momentum phase to obtain the transformed hodograph H'
\end{itemize}
With this procedure we will get the finite transformation corresponding to our symmetry and not just the infinitesimal one like in the previous section. In particular let us see that the finite transformation generate by $A_1$ corresponds to a rotation by the angle $\psi$ in the $Q_0,Q_1$ 2-plane (and similarly for other components of ${\vec A}$). We have
\be
 P=(0,p_i) \in H \quad \to \quad Q=\left( \frac{p^2-p_0^2}{p^2+p_0^2}p_0,\frac{2p_0^2p_i}{p^2+p_0^2} \right) \in G
\ee
\begin{align}
Q'= \begin{bmatrix} \cos \psi & -\sin \psi & 0 & 0 \\  \sin \psi & \cos \psi & 0 & 0 \\ 0 & 0 & 1 & 0 \\ 0 & 0 & 0 & 1 \\ \end{bmatrix} \times Q
= \left( \begin{array}{c} Q_0 \cos \psi - Q_1 \sin \psi  \\ Q_0 \sin \psi + Q_1 \cos \psi \\ Q_2 \\ Q_3  \end{array} \right) \in G'
\end{align}
\be
Q' \in G' \quad \to \quad P'= \left(0,\frac{p_0}{p_0-Q'_0} Q'_i \right) \in H'
\ee
that is
\begin{align}
P'_1 &=\frac{p_0}{p_0-Q'_0}Q'_1=\frac{p_0(Q_0 \sin \psi + Q_1 \cos \psi )}{p_0-Q_0 \cos \psi + Q_1 \sin \psi} \nonumber\\
&= \dfrac{p_0 \left((p^2-p_0^2)p_0 \sin \psi + 2p_0^2 p_1 \cos \psi \right)}{(p^2+p_0^2)p_0 -(p^2-p_0^2)p_0 \cos \psi + 2p_0^2 p_1 \sin \psi}~, \label {conformalp1}\\
 \nonumber \\
P'_{a} &=  \frac{2p_0^3p_{a}}{(p^2+p_0^2)p_0 -(p^2-p_0^2)p_0 \cos \psi + (2p_0^2)p_1 \sin \psi} \quad \text{for   } a=2,3~. \label {conformalpmu}
\end{align}
\\

We shall now interpret the hypersphere rotations directly within ordinary momentum space  in terms of conformal transformations. More precisely, we shall prove that this transformation consists of a composition of a special conformal transformation, a translation and a scaling (or dilation). We know that the conformal group, the group of symmetries that conserves angles or correspond to a scaling of the metric tensor, $g_{\mu\nu}(x) \to \Omega(x) g_{\mu\nu}$, is composed of 4 kinds of symmetries
\begin{align}
&\text{translation:} &p'_{\mu}&=p_{\mu}+a_{\mu}\\
&\text{rotation:} &p'_{\mu}&=M_{\mu}^{\nu}p_{\nu}\\
&\text{special conformal transformation:} \qquad &p'_{\mu}&= \dfrac{p_{\mu} -b_{\mu}p^2}{1-2(\vec{b}\cdot\vec{p}+b^2p^2)}\\
&\text{scaling:}  &p'_{\mu}&= \alpha p_{\mu}
\end{align}
Thus a composition of special conformal transformation, translation and scaling of transformation can  be written
\be
p'_{\mu} = \alpha~ \dfrac{p_{\mu} -b_{\mu}p^2+a_{\mu}(1-2\vec{b}\cdot\vec{p}+b^2p^2)}{1-2\vec{b}\cdot\vec{p}+b^2p^2}~,
\ee
Since in our example, we only transformed the $1$-axis, $\vec{b}=b\hat{\vec{x}}$, $\vec{a}=a\hat{\vec{x}}$ we search a transformation that will looks like 
\begin{align}
p'_1 &= \alpha~ \dfrac{p_1 -bp^2+a(1-2bp_1+b^2p^2)}{1-2bp_1+b^2p^2}  \label{tansformation1}~,\\
p'_{a} &= \alpha~ \dfrac{p_{a}}{1-2bp_1+b^2p^2} \quad \text{for}~a=2,3 \label{tansformationmu}~.
\end{align}
With simple computation we can indeed rewrite (\ref{conformalpmu})
\be
P'_{a} = \left(\frac{2}{1+\cos \psi}\right)~ \frac{ p_{\mu}}{1+2\dfrac{\sin \psi}{1+\cos \psi}~\dfrac{p_1}{p_0}+\dfrac{1-\cos \psi}{1+\cos \psi}~\dfrac{p_1^2}{p_0^2}} \quad \text{for   } a=2,3~,
\ee
which gives 
\be
b=-\frac{\sin \psi}{1+\cos \psi},~~\alpha = \frac{2}{1+\cos \psi}~.
\ee
and we can rewrite (\ref{conformalp1})
\be
P'_1 = \left(\frac{2}{1+\cos \psi}\right)~ \dfrac{\cos \psi~ p_1 +\frac{1}{2} \sin \psi~ \left(\dfrac{p_1^2}{p_0^2}-1\right)}{1+2\dfrac{\sin \psi}{1+\cos \psi}~\dfrac{p_1}{p_0}+\dfrac{1-\cos \psi}{1+\cos \psi}~\dfrac{p_1^2}{p_0^2}}~,
\ee
which gives 
\be
a=-\frac{1}{2}\sin \psi~.
\ee
We thus proved that the transformation corresponding to the LRL vector is a conformal transformation between orbits of different eccentricity but same energy, composed of a translation, special conformal transformation and scaling.

Summarizing this section we saw that $\vec{A}$ can be interpreted to generate rotations of Fock's hypersphere, which translate back to conformal transformations in ordinary momentum space which conserve $p_0$, and we gained some intuition for some of steps in the argument.

\part{Relativistic anomalies}
\label{part:anomalies}

Having studied  in the first part  the symmetries of the non relativistic problem and the corresponding conserved quantities, we now want to focus on anomalies (that is broken symmetries or unconserved physical quantities) arising when we add perturbations to the non relativistic Hamiltonian to take into account several General Relativity effects.\\
We will study 3 cases : 1PN approximation, spin-orbit coupling and spin-spin coupling.

\section{1PN Post-Newtonian Approximation}
\label{sec:H1PNr}

At 1PN approximation, the elliptical orbits will not be closed anymore and will precess. At this order, $\vec{A}$ is no more conserved. We will compute $\langle{\dot{\vec{A}}}\rangle$ using our general method to obtain the anomaly and the angular velocity of precession in a rather elegant and simple manner.
The 1PN Lagrangian is given by \cite{LorentzDroste17,Damour1985} and references therein
\be
{\cal L} =L_0+\frac{1}{c^2}L_{1PN} \label{Lagrangian1PN}~,
\ee
with
\be
L_0=\frac{1}{2}m_1\vec{v_1}^2+\frac{1}{2}m_2\vec{v_2}^2+G\frac{m_1m_2}{r}~,
 \ee
 \begin{align}
 L_{1PN} &=\frac{1}{8}m_1\vec{v_1}^4+\frac{1}{8}m_2\vec{v_2}^4+G\frac{m_1m_2}{2r}\left[3\vec{v_1}^2+3\vec{v_2}^2-8\vec{v_1}\cdot\vec{v_2} \right. \nonumber \\
 &~\left. +(\vec{v_1}\cdot\vec{v_2}-(\hat{\vec{r}}\cdot\vec{v_1})(\hat{\vec{r}}\cdot\vec{v_2}))\right]-G^2\frac{m_1m_2(m_1+m_2)}{2r^2}~.
 \end{align}
The interpretation of the $\frac{1}{c^2}$ correction terms to the Newtonian Lagrangian is quite clear using the EFT description \cite{Kol2007}: the first two terms represents the correction to the kinetic energies, the next two terms the gravitation of kinetic energy, the following $8\vec{v_1}\cdot\vec{v_2}$ term represents the gravitomagnetic current-current interaction (a moving charge creates a gravitomagnetic field that will interact with other moving charge), the next term represents the retardation effect due to the finite speed of light and finally the last term represents the contribution of potential energy to the gravitational interaction.

Invariance of (\ref{Lagrangian1PN}) under spatial translations and Lorentz boosts implies, due to Noether's theorem, the conservation of the total linear momentum of the system and of the relativistic center of mass. Then the relative Lagrangian can be found in \cite{Damour1985} and the relative Hamiltonian is given by \cite{Damour1988} (with reduced $\vec{p}$)
\begin{align} 
H &=  H_0 + \frac{1}{c^2}H_{1PN}\\
&= \left[ \frac{\vec{p}^2}{2} - \frac{\alpha}{r} \right]+\frac{1}{c^2}\left[\frac{(3\nu-1)}{8}\vec{p}^4-\frac{\alpha(3+\nu)}{2}\frac{\vec{p}^2}{r}-\frac{\alpha\nu} {2}\frac{(\vec{p}\cdot\hat{\vec{r}})^2}{r}+\frac{\alpha^2}{2r^2}\right]~.
\end{align} 
with $\nu=\frac{\mu}{m}=\frac{m_1 m_2}{(m_1 + m_2)^2}$ and $\alpha = Gm = G (m_1 + m_2)$.

\subsection{$\langle H_{1PN} \rangle$  and $\langle\dot{\vec{A}}\rangle_{1PN}$}

We shall compute $\langle\dot{\vec{A}}\rangle$ from (\ref{Adotmethod}) and for that purpose we shall compute now $\langle H_{1PN} \rangle$ where the averaging is performed over Newtonian orbit.

\subsubsection*{$H_{1PN}$}

Let us write\footnote{
The following definition of $A$ is used only within this subsection and should not be confused with the LRL vector.}
\begin{equation}
H_{1PN}=Ap^4+B\frac{p^2}{r}+C\frac{(\vec{p}.\vec{\hat{r}})^2}{r^3}+D\frac{1}{r^2}~, \label{H1pnabcd}
\end{equation}

\[
\text{with}~\left\{
\begin{array}{ll}
A =\frac{(3\nu-1)}{8}\\
B = -\frac{\alpha(3+\nu)}{2}\\
C=-\frac{\alpha\nu}{2}\\
D= \frac{\alpha^2}{2}
\end{array}
\right.
\]
The averaging is carried over Newtonian orbits according to the averaging method, see section \ref{sec:method}. For these orbits $(\ref{E}-\ref{rdot})$ we can express $p$ in terms of $r$
\begin{align}
p^2 &= 2(E+\frac{\alpha}{r})~,\\
 p^4 &= 4(E^2+2E\frac{\alpha}{r}+\frac{\alpha^2}{r^2})~,\\
\frac{(\vec{p}\cdot\hat{\vec{r}})^2}{r} &= \frac{p_{r}^2}{r}=\frac{p^2}{r}-\frac{p_{\theta}^2}{r}=\frac{2}{r}(E+\frac{\alpha}{r})-\frac{L^2}{r^3}~.
\end{align}
\\
substituting back into (\ref{H1pnabcd}) we can express $H_{1PN}$ as a function of the conserved quantities $E$ and $L$, and in terms of $r(t)$
\begin{equation} \label {eqn:H1pn}
H_{1PN}=4AE^2+\frac{1}{r}[8A\alpha+2(B+C)]E+\frac{1}{r^2}[4A\alpha^2+2 \alpha(B+C)+D]-\frac{1}{r^3}CL^2~.
\end{equation}

\subsubsection*{Time averaging}

The average of a physical quantities Q over a complete orbit is given by
\[
\langle Q \rangle = \frac{1}{T} \oint dt Q~,
\]
Using again  $(\ref{E}-\ref{rdot})$ we can transform the integration from $t$ to $\theta$
\begin{align}
\nonumber\langle Q \rangle &= \frac{1}{T} \oint dt Q = \frac{1}{T} \oint \frac{d\theta}{\dot{\theta}} Q\\ 
\non &=\frac{1}{T} \frac{1}{L}\oint d\theta r^2 Q = \frac{1}{2\pi} \left(\frac{\alpha}{a^3}\right)^{\frac{1}{2}} \frac{1}{L}\oint d\theta r^2 Q\\
&= \frac{1}{2\pi} \frac{(-2E)^{\frac{3}{2}}}{\alpha L}\oint d\theta r(\theta)^2 Q(r,\theta)~. \label {Qaverage}
\end{align}
The integration can be performed through analytic continuation and the residue method (see Appendix \ref{app:Residu}) :
\begin{align}
\langle\frac{1}{r}\rangle &=  \frac{1}{2\pi} \frac{(-2E)^{\frac{3}{2}}}{\alpha L}\oint d\theta r(\theta) = \frac{1}{2\pi} \frac{(-2E)^{\frac{3}{2}}}{\alpha L} \frac{L^2}{\alpha} \oint d\theta (1+e\cos\theta)^{-1} \non \\
&=  \frac{1}{2\pi} \frac{(-2E)^{\frac{3}{2}}L}{\alpha^2} \frac{2\pi}{(1-e^2)^ \frac{1}{2}}= \frac{(-2E)^{\frac{3}{2}}L}{\alpha^2} \frac{\alpha}{L (-2E)^ \frac{1}{2}}=\frac{(-2E)}{\alpha}~. \label {eqn:r1m}
\end{align}
which is equivalent to the Virial Theorem $E=\frac{1}{2}\langle\frac{\alpha}{r}\rangle$.
In fact some of the integrals are elementary
\begin{align}
\langle\frac{1}{r^2}\rangle &=  \frac{1}{2\pi} \frac{(-2E)^{\frac{3}{2}}}{\alpha L}\oint d\theta  = \frac{(-2E)^{\frac{3}{2}}}{\alpha L}~. \label {eqn:r2m}
\end{align}
and
\begin{align}
\langle\frac{1}{r^3}\rangle &=  \frac{1}{2\pi} \frac{(-2E)^{\frac{3}{2}}}{\alpha L}\oint d\theta r(\theta)^-1 = \frac{1}{2\pi} \frac{(-2E)^{\frac{3}{2}}}{\alpha L} \frac{\alpha}{L^2}\oint d\theta (1+e \cos\theta) \nonumber\\
&=\frac{1}{2\pi} \frac{(-2E)^{\frac{3}{2}}}{L^3} 2\pi = \frac{(-2E)^{\frac{3}{2}}}{L^3}~. \label {eqn:r3m}
\end{align}
\\
Inserting $(\ref {eqn:r1m})(\ref {eqn:r2m})(\ref {eqn:r3m})$  in equation $(\ref {eqn:H1pn})$ we get
\begin{align} \label {eqn:H1pnm}
\langle H_{1PN} \rangle &= 4AE^2+[8A\alpha+2(B+C)]E \frac{(-2E)}{\alpha}\nonumber\\
&+[4A\alpha^2+2\alpha(B+C)+D]\frac{(-2E)^{\frac{3}{2}}}{\alpha L} - CL^2 \frac{(-2E)^{\frac{3}{2}}}{L^3}~,
\end{align}\\
substituting for A,B,C,D we obtain
\begin{equation}
\langle H_{1PN} \rangle = \frac{15-\nu}{2}E^2-3 \frac{\alpha(-2E)^{\frac{3}{2}}}{L}~.
\end{equation}
 
\subsubsection*{$\langle\dot{\vec{A}}\rangle_{1PN}$}
Recalling that $\vec{A}$ and E commute at Newtonian order, we have
\begin{equation}
\langle\dot{\vec{A}}\rangle_{1PN} = \{\vec{A},\langle H_{1PN} \rangle\}=-3\alpha(-2E)^{\frac{3}{2}} \{\vec{A},\frac{1}{L}\}
\end{equation}
Now, we know that since $\vec{A}$ is a vector
\[
\{A^i,L^j\}=\epsilon_{ijk}A^k\\
\]
and given the Leibniz rules
\[
\{A^i,L^{j1}L^{j2}\}=\{A^i,L^{j1}\}L^{j2}+L^{j1}\{A^i,L^{j2}\}\\
\]
we can infer that for a given analytic function f(L) 
\begin{equation} \label {eqn:Apoisson}
 \{A^i,f(L)\}=\epsilon_{ijk} \frac{\partial f(L)}{\partial L^j}A^k
\end{equation}
since we can expand f(L) in Taylor series as a product of ~$L^j$ and use Leibniz rules. Thus we reach the main result of this section
\begin{equation}
\langle\dot{\vec{A}}\rangle_{1PN} =-3\alpha(-2E)^{\frac{3}{2}} \{\vec{A},\frac{1}{L}\} = 3 \frac{\alpha(-2E)^{\frac{3}{2}}}{L^3} (\vec{L} \times \vec{A})~. \label{A1pn}
\end{equation}
\\
The LRL vector $\vec{A}$ is thus rotating with an angular velocity
\begin{equation} \label{angularvelocityH1pn}
\vec{\Omega}_{A_{1PN}} = 3 \frac{\alpha}{a^3 (1-e^2)^{\frac{3}{2}}} \vec{L} \equiv \frac{3\alpha}{b^3}~\vec{L}~.
\end{equation}

The angular velocity is around the $\hat{\vec{ L}}$ axis, so the plane orbit remains the same (this is of course a confirmation that at this order $\vec{L}$ is a conserved quantity) but the orbit is precessing (since the LRL vector $\vec{A}$ which gives the direction of the periapsis is now rotating).

In $(\ref{angularvelocityH1pn})$ we reproduced a well-known expression but in a simpler and quicker way than \cite{Barker1970} that mention``after a rather lengthy calculation''...while avoiding Hamilton-Jacobi Equation, angle-action and canonical perturbation approach as in \cite{Goldstein1,BinneyTremaine}.
\subsection{Conserved quantities}
As in the newtonian case, E and $\vec{L}$ are conserved quantities.\\
$\vec{A}$ is no more a constant of motion, buts  its magnitude $A^2$ remains, since the vector is just rotating (\ref{A1pn}).

\section{ Spin-orbit coupling, induced apsidal motion and orbit precession}
 \label{sec:Hso}

For spinning objects the Newtonian equations of motion are supplemented by
\be
\dot{\vec{S}}_a=0 \quad a=1,2
\ee

Hence $\vec{S}_a$ are conserved. However relativistic corrections couple the spin to the motion as described by the spin-orbit Hamiltonian (at leading order) \cite{Barker1970}
\begin{equation*} \label {eqn:Hso}
H_{SO}=\frac{G}{c^2 r^3} [ (2+\frac{3m_2}{2m_1})(\vec{L}\cdot \vec{S_1})+ (2+\frac{3m_1}{2m_2})(\vec{L}\cdot \vec{S_2})]~.
\end{equation*}

This correction is due to a gravitomagnetic dipole - current interaction (a spinning object is equivalent to a gravitomagnetic dipole which creates a field that will interact with a moving charge). The order of this effect $1.5PN+1a^*$ where $a^*:=CGS/(GM)^2$ is a dimensionless parameter which quantifies the size of the spin (assuming $S_1\sim S_2$, $m_1\sim m_2$).\footnote{By  $1.5PN+1a^*$ we mean that $H_{SO}$ is of order $(\frac{v^2}{c^2})^{1.5} (a^*)^1 H_0$} For black holes $|a^*|\le 1$ and hence in this case the order cannot be lower than $1.5PN$.

The spin-orbit coupling breaks $\vec{L}$, $\vec{S_a}$ and  $\vec{A}$ and causes all of them to precess, as we shall see using our method.

We can write
\begin{equation}
H_{SO} = \frac{\vec{L}\cdot \vec{S}}{r^3}~,
\end{equation}
with
\be
\vec{S} = \frac{G}{c^2} [ (2+\frac{3m_2}{2m_1})\vec{S_1}+ (2+\frac{3m_1}{2m_2})\vec{S_2}) ]~.
\ee

\subsection{$\langle H_{SO} \rangle$}

Using $(\ref {eqn:r3m})$ the computation of $\langle H_{SO} \rangle$ in terms of the conserved quantities space E and $\vec{L}$ is immediate
\begin{equation*}
\langle  H_{SO} \rangle = \langle \frac{1}{r^3} \rangle \vec{L}\cdot \vec{S} = \frac{(-2E)^{\frac{3}{2}}}{L^3} \vec{L}\cdot \vec{S}~.
\end{equation*}

\subsection{$\langle \dot{\vec{L}} \rangle_{SO}$,  $\langle \dot{\vec{S}}_1 \rangle_{SO}$ and $\langle \dot{\vec{S}}_2 \rangle_{SO}$ }

\begin{figure}[!hb]
\includegraphics[width=4cm]{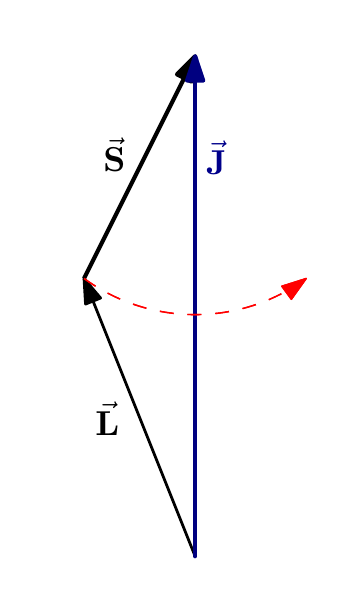}\centering
\caption{$\vec{L}$ and $\vec{S}$ are rotating around $\vec{J}$ the total angular momentum of the system.}\label{fig:spinorbitrotation}
\end{figure}

First of all, since $\vec{S}_1, \vec{S}_2$ and $\vec{L}$ belong to different vector spaces, we have the following Poisson bracket rules 
\ben
\{\vec{S}_1,\vec{S}_2\}=\{\vec{S}_1,\vec{L}\}=\{\vec{S}_1,\vec{L}\}=0
\een
We have
\begin{align*}
\langle \dot{\vec{L}} \rangle_{SO} &= \{\vec{L},\langle H_{SO} \rangle\}\\
\langle\dot{L_i}\rangle _{SO} &= \{L_i,\frac{(-2E)^{\frac{3}{2}}}{L^3} L_j S_j\}\\
&=(-2E)^{\frac{3}{2}} [\frac{S_j}{L^3}\{L_i,L_j\}+S_j L_j \{L_i,\frac{1}{L^3}\}+\frac{L_j}{L^3}\{L_i,S_j \}]\\
&= (-2E)^{\frac{3}{2}} [\frac{S_j}{L^3}\{L_i,L_j\}] = \frac{ (-2E)^{\frac{3}{2}}}{L^3} \epsilon_{ijk}S_jL_k~,
\end{align*}\\
 that is
\be
\langle \dot{\vec{L}} \rangle_{SO} = \frac{ (-2E)^{\frac{3}{2}}}{L^3} \vec{S} \times \vec{L} =  \frac{ (-2E)^{\frac{3}{2}}}{L^3} \frac{G}{c^2} [ (2+\frac{3m_2}{2m_1})\vec{S_1}+ (2+\frac{3m_1}{2m_2})\vec{S_2}) ] \times \vec{L}~. \label{Ldot}
\ee
We have
\begin{align*}
\langle \dot{\vec{S}}_1 \rangle_{SO} &= \{\vec{S_1},\langle H_{SO} \rangle\}\\
\langle\dot{S}_{1i}\rangle _{SO} &= \{S_{1i},\frac{(-2E)^{\frac{3}{2}}}{L^3} L_j S_j\}\\
&= (-2E)^{\frac{3}{2}} [\frac{L_j}{L^3}\frac{G}{c^2}(2+\frac{3m_2}{2m_1})\{S_{1i},S_{1j}\}]\\
&= \frac{ (-2E)^{\frac{3}{2}}}{L^3}\frac{G}{c^2}(2+\frac{3m_2}{2m_1})\epsilon_{ijk}L_jS_{1k}~,
\end{align*}\\
 that is
\begin{equation}
\langle \dot{\vec{S}}_1 \rangle_{SO} = \frac{ (-2E)^{\frac{3}{2}}}{L^3}\frac{G}{c^2}(2+\frac{3m_2}{2m_1}) \vec{L} \times \vec{S_1}~, \label{S1dot}
\end{equation}
and similarly
\begin{equation}
\langle \dot{\vec{S}}_2 \rangle_{SO} = \frac{ (-2E)^{\frac{3}{2}}}{L^3}\frac{G}{c^2}(2+\frac{3m_1}{2m_2})\vec{L} \times \vec{S_2}~. \label{S2dot}
\end{equation}
\\
We thus see that we have $\dot{\vec{L}} = - \dot{\vec{S}}_1 - \dot{\vec{S}}_2$, that is $ \vec{J} =\vec{L} +\vec{S}$ the total angular momentum of the system is a conserved quantity, and we can rewrite
\begin{align}
\langle \dot{\vec{L}} \rangle_{SO} &= \frac{ (-2E)^{\frac{3}{2}}}{L^3} \vec{S} \times \vec{L} = \frac{ (-2E)^{\frac{3}{2}}}{L^3} \vec{(L+S)} \times \vec{L}=\frac{ (-2E)^{\frac{3}{2}}}{L^3} \vec{J} \times \vec{L}~,\\
\langle \dot{\vec{S}} \rangle_{SO} &= \frac{ (-2E)^{\frac{3}{2}}}{L^3} \vec{L} \times \vec{S} = \frac{ (-2E)^{\frac{3}{2}}}{L^3} \vec{(L+S)} \times \vec{S}=\frac{ (-2E)^{\frac{3}{2}}}{L^3} \vec{J} \times \vec{S}~.
\end{align}
Thereby, spin-orbit coupling induce a rotation of $\vec{L}, \vec{S}$ around $\vec{J}$ total angular momentum of the system, as shown in figure  \ref{fig:spinorbitrotation}, with the same angular velocity 
\be
\Omega_{\vec{L}_{SO}} = \Omega_{\vec{S}_{SO}} =  \frac{ (-2E)^{\frac{3}{2}}}{L^3} \vec{J} \label{omegaLso}~.
\ee
This means that the orbital plane (always perpendicular to $\vec{L}$) is not fixed anymore but it rotates around $\vec{J}$. This orbital plane rotation is called the apsidal motion.

\subsection{$\langle \dot{\vec{A}} \rangle_{SO} $ }

Recalling $(\ref {eqn:Apoisson})$ we have
\begin{align*}
\langle \dot{\vec{A}} \rangle_{SO} &= \{\vec{A},\langle H_{SO} \rangle\}\\
\langle\dot{A_i}\rangle _{SO} &= \{A_i,\frac{(-2E)^{\frac{3}{2}}}{L^3} L_n S_n\}= (-2E)^{\frac{3}{2}}  \{A_i,\frac{ L_n S_n}{L^3}\}\\
&=  (-2E)^{\frac{3}{2}} \epsilon_{ijk} \frac{\partial}{\partial L^j} (\frac{ L_n S_n}{L^3})A^k\\
&= (-2E)^{\frac{3}{2}} \epsilon_{ijk}[\frac{1}{L^3}\frac{\partial}{\partial L^j}( L_n S_n)+ (L_n S_n)\frac{\partial}{\partial L^j}(L_pL_p)^{-\frac{3}{2}}]A^k\\
&=(-2E)^{\frac{3}{2}} \epsilon_{ijk}[\frac{1}{L^3} S_n \delta_{nj} + (L_n S_n) (-\frac{3}{2})(L_pL_p)^{-\frac{5}{2}} 2 \delta_{pj} L_p]A^k\\
&=(-2E)^{\frac{3}{2}} \epsilon_{ijk}[\frac{1}{L^3} S_j -3 \frac{(L_n S_n)}{L^5} L_j ]A^k~,
\end{align*}\\
and finally
\begin{equation}
\langle\dot{\vec{A}}\rangle _{SO} = \frac{(-2E)^{\frac{3}{2}}}{L^3} (\vec{S}-3 \frac{(\vec{L}\cdot\vec{S})}{L^2} \vec{L})\times\vec{A}~.
\label{SO-Adot}
\end{equation}

Thus, the spin-orbit coupling is adding another contribution to the LRL vector $\vec{A}$ rotation, with an angular velocity 
\begin{equation}
\vec{\Omega}_{A_{SO}} = \frac{(-2E)^{\frac{3}{2}}}{L^3} (\vec{S}-3 \frac{(\vec{L}\cdot\vec{S})}{L^2} \vec{L}) \label{Aso}~.
\end{equation}\\
Altogether, the anomalies given in $(\ref{omegaLso}, \ref{Aso})$ reproduce the corresponding expressions in \cite{Barker1970} in the test mass limit and more generally in \cite{Damour1988}. Actually, Damour and Sch\"{a}fer in  \cite{Damour1988}, use also a similar method involving vectorial and Poisson brackets calculus, but they average  $\dot{\vec{A}}_{SO}$ only after the computation of $\{\vec{A},H_{SO}\}$. 

To determine the contribution of the spin-orbit coupling to the precession of the periapsis within the orbit plane, we just have to calculate the component of the angular velocity along the $\hat{\vec{ L}}$ axis, that is $\vec{\Omega}_{A_{SO}} \cdot \hat{\vec{ L}}$.
\begin{equation}
\vec{\Omega}_{A_{SO}}\cdot \hat{\vec{ L}} = \frac{(-2E)^{\frac{3}{2}}}{L^3} (\vec{S} \cdot \vec{\hat{L}}-3\frac{L(\vec{S} \cdot \vec{\hat{L}})}{L^2} L) = (-2)  \frac{(-2E)^{\frac{3}{2}}}{L^3} (\vec{S} \cdot \vec{\hat{L}})~.
\end{equation}\\
It depends only on the orientation of $\vec{S}$ and should be compared with the 1PN anomaly (\ref{angularvelocityH1pn}). 

\subsection{Conserved quantities}
H the Hamiltonian is conserved by construction so E is a constant of motion.\\
From (\ref{Ldot}, \ref{S1dot}, \ref{S2dot}, \ref{Aso}) we see that
\begin{itemize}
\item if $\vec{S}$, the total spin of the system, is parallel (or antiparallel) to $\vec{L}$, the orbital angular momentum of the system, then $\vec{L}$ and $\vec{S}$ are also constant of motion,
\item for general orientation of $\vec{S}$, $\vec{L}$ and $\vec{S}$ are not conserved but $ \vec{J} =\vec{L} +\vec{S}$  the total angular momentum of the system is conserved,
\item the magnitude of $\vec{L}$ and $\vec{S}$ are always constant since the vectors are just rotating and the angle between $\vec{L}$ and $\vec{S}$ is also a constant since $\frac{d}{dt}(\vec{L}\cdot\vec{S})=0$,
\item $\vec{A}$ the LRL vector is not conserved, whatever the general orientation of $\vec{S}$, but its magnitude is constant since the vector is just rotating. 
\end{itemize}

\section{ Spin-Spin coupling}
 \label{sec:Hss}

Spinning objects are also subject to the spin-spin interaction given by the perturbation Hamiltonian \cite{Barker1970}
\be \label {Hss}
H_{SS}=\frac{G}{c^2 r^3} [3(\vec{S_1}\cdot \vec{\hat{r}})(\vec{S_2}\cdot \vec{\hat{r}})-(\vec{S_1}\cdot \vec{S_2})]~.
\ee
This relativistic correction is due to a gravitomagnetic dipole - dipole interaction (a dipole in a gravitomagnetic field is experiencing a torque that tries to anti-align its intrinsic angular momentum with the gravitomagnetic field). The order of this effect is $2PN+1a_1^*+1a_2^*$.

\subsection{$\langle H_{SS} \rangle$}

The time average of the second term of ($\ref {Hss}$) is easily computed using $(\ref {eqn:r3m})$
\be
\langle \frac{1}{r^3}(\vec{S_1}\cdot \vec{S_2})\rangle = \frac{(-2E)^{\frac{3}{2}}}{L^3} (\vec{S_1}\cdot \vec{S_2})~.
\ee
For the first term, it is more complex, we need to use $(\ref {Qaverage})$  and find the right expression for $\vec{\hat{r}}$. For that we write $\vec{\hat{r}}$ in the orthonormal referential $(\vec{\hat{A}},\vec{\hat{L}}\times \vec{\hat{A}},\vec{\hat{L}})$. Indeed we have
\be
\vec{\hat{r}} = \cos\theta \vec{\hat{A}} + \sin\theta (\vec{\hat{L}}\times \vec{\hat{A}}) + 0 \vec{\hat{L}}~,
\ee
since at Newtonian order, $\vec{\hat{A}}$ gives the direction of the semi-major axis and is in the orbital plane (\ref{A1}).
We can then compute
\begin{align*}
\langle\frac{3}{r^3} [3(\vec{S_1}\cdot \vec{\hat{r}})(\vec{S_2}\cdot \vec{\hat{r}})] \rangle = & 
\frac{3}{2\pi} \frac{(-2E)^{\frac{3}{2}}}{L^3} \oint d\theta (1+e \cos\theta)\\
& [(\vec{S_1}\cdot \vec{\hat{A}})\cos\theta+(\vec{S_1}\cdot \vec{\hat{L}}\times \vec{\hat{A}}) \sin\theta]\\
& [(\vec{S_2}\cdot \vec{\hat{A}})\cos\theta+(\vec{S_2}\cdot \vec{\hat{L}}\times \vec{\hat{A}}) \sin\theta]\\
=& \frac{3}{2\pi} \frac{(-2E)^{\frac{3}{2}}}{L^3} \oint d\theta (1+e \cos\theta)\\
&[(\vec{S_1}\cdot \vec{\hat{A}})(\vec{S_2}\cdot \vec{\hat{A}})\cos^2 \theta+(\vec{S_1}\cdot\vec{\hat{A}}) (\vec{S_2}\cdot \vec{\hat{L}}\times \vec{\hat{A}})\cos\theta\sin\theta\\
&+ (\vec{S_2}\cdot\vec{\hat{A}})(\vec{S_1}\cdot \vec{\hat{L}}\times \vec{\hat{A}})\cos\theta\sin\theta\\
&+ (\vec{S_1}\cdot \vec{\hat{L}}\times \vec{\hat{A}})(\vec{S_2}\cdot \vec{\hat{L}}\times \vec{\hat{A}})\sin^2\theta]~,
\end{align*}
we get
\begin{align*}
\langle\frac{3}{r^3} [3(\vec{S_1}\cdot \vec{\hat{r}})(\vec{S_2}\cdot \vec{\hat{r}})] \rangle &=
\frac{3}{2\pi} \frac{(-2E)^{\frac{3}{2}}}{L^3}[(\vec{S_1}\cdot \vec{\hat{A}})(\vec{S_2}\cdot \vec{\hat{A}})\pi
+(\vec{S_1}\cdot \vec{\hat{L}}\times \vec{\hat{A}})(\vec{S_2}\cdot \vec{\hat{L}}\times \vec{\hat{A}})\pi]\\
&= \frac{3}{2} \frac{(-2E)^{\frac{3}{2}}}{L^3}[(\vec{S_1}\cdot \vec{S_2} - (\vec{S_1}\cdot \vec{\hat{L}})(\vec{S_2}\cdot \vec{\hat{L}})]~,
\end{align*}
and finally
\be
\langle H_{SS} \rangle = \frac{(-2E)^{\frac{3}{2}}}{2L^3}[(\vec{S_1}\cdot \vec{S_2}) -3 (\vec{S_1}\cdot \vec{\hat{L}})(\vec{S_2}\cdot \vec{\hat{L}})]~.\\
\ee

\subsection{$\langle \dot{\vec{L}} \rangle_{SS}$,  $\langle \dot{\vec{S}}_1 \rangle_{SS}$ and $\langle \dot{\vec{S}}_2 \rangle_{SS}$ }

We have
\begin{align*}
\langle \dot{\vec{L}}\rangle_{SS} &= \{\vec{L},\langle H_{SS} \rangle\}\\
\langle\dot{L}_{i}\rangle _{SS} &= -3\frac{(-2E)^{\frac{3}{2}}}{2L^3} \{L_{i},(S_{1l}\hat{L}_{l})(S_{2m}\hat{L}_{m})\}\\
&= -3\frac{(-2E)^{\frac{3}{2}}}{2L^5} [(S_{1l}L_{l})S_{2m}\{L_{i},L_{m})\}
+(S_{2m}L_{m})S_{1l}\{L_{i},L_{l})\}\\
&=  -3\frac{(-2E)^{\frac{3}{2}}}{2L^5} [(S_{1l}L_{l})S_{2m} \epsilon_{imk}L_{k}+ (S_{2m}L_{m})S_{1l} \epsilon_{ila}L_{a}]~,
\end{align*}
That is
\be
\langle \dot{\vec{L}} \rangle_{SS} = -3\frac{(-2E)^{\frac{3}{2}}}{2L^3} [(\vec{S_1}\cdot \vec{\hat{L}})\vec{S_2}+ (\vec{S_2}\cdot \vec{\hat{L}})\vec{S_1}] \times \vec{L}~. \label{Ldotss}
\ee
We have
\begin{align*}
\langle \dot{\vec{S}}_1 \rangle_{SS} &= \{\vec{S_1},\langle H_{SS} \rangle\}\\
\langle\dot{S}_{1i}\rangle _{SS} &= \frac{(-2E)^{\frac{3}{2}}}{2L^3} [\{S_{1i},S_{1j}S_{2j}\}-3 \{S_{1i},(S_{1l}\hat{L}_{l})(S_{2m}\hat{L}_{m})\}]\\
&=  \frac{(-2E)^{\frac{3}{2}}}{2L^3} [\epsilon_{ijk}S_{1k}S_{2j}-3\epsilon_{iln}S_{1n}\hat{L}_{l} (S_{2m}\hat{L}_{m})~,
\end{align*}
That is
\be
\langle \dot{\vec{S}}_1 \rangle_{SS} = \frac{(-2E)^{\frac{3}{2}}}{2L^3} [\vec{S_2}-3(\vec{S_2}\cdot \vec{\hat{L}})\vec{\hat{L}}] \times \vec{S}_1~,\label{S1dotss}
\ee
and similarly 
\be
\langle \dot{\vec{S}}_2 \rangle_{SS} = \frac{(-2E)^{\frac{3}{2}}}{2L^3} [\vec{S_1}-3(\vec{S_1}\cdot \vec{\hat{L}})\vec{\hat{L}}] \times \vec{S}_2~. \label{S2dotss}
\ee
\\
Thereby, the spin-spin coupling induces a precession of $\vec{L}, \vec{S}_1, \vec{S}_2,$ with angular velocities
\begin{align}
\Omega_{\vec{L}_{SS}} &= -3\frac{(-2E)^{\frac{3}{2}}}{2L^3} [(\vec{S_1}\cdot \vec{\hat{L}})\vec{S_2}+ (\vec{S_2}\cdot \vec{\hat{L}})\vec{S_1}]~,\label{omegalss}\\
\Omega_{\vec{S}_{1SS}} &= -3\frac{(-2E)^{\frac{3}{2}}}{2L^3}  [\vec{S_2}-3(\vec{S_2}\cdot \vec{\hat{L}})\vec{\hat{L}}]~, \label{omegaS1ss}\\
\Omega_{\vec{S}_{2SS}} &= \frac{(-2E)^{\frac{3}{2}}}{2L^3} [\vec{S_1}-3(\vec{S_1}\cdot \vec{\hat{L}})\vec{\hat{L}}]~. \label{omegaS2ss}
\end{align}

\subsection{$\langle \dot{\vec{A}} \rangle_{SS} $ }

\begin{align}
\langle \dot{\vec{A}} \rangle_{SS} &= \{\vec{A},\langle H_{SS} \rangle\} \non\\
&= \frac{ (-2E)^{\frac{3}{2}}}{2} \{\vec{A},  [\frac{(\vec{S_1}\cdot \vec{S_2})}{L^3}-3 \frac{(\vec{S_1}\cdot \vec{L})(\vec{S_2}\cdot \vec{L})}{L^5}] \}~. \label{Ass}
\end{align}
Using again $(\ref {eqn:Apoisson})$ we have
\begin{align}
\{A_i, \frac{(\vec{S_1}\cdot \vec{S_2})}{L^3} \} &= \epsilon_{ijk} \frac{\partial}{\partial L^j} (\frac{(\vec{S_1}\cdot \vec{S_2})}{L^3})A^k \non\\
&= -3 (\vec{S_1}\cdot \vec{S_2})\epsilon_{ijk}\frac{L_j A^k}{L^5}~, \label{Ass1}
\end{align}
and
\begin{align}
\{A_i, \frac{(\vec{S_1}\cdot \vec{L})(\vec{S_2}\cdot \vec{L})}{L^5} \} &= \epsilon_{ijk} \frac{\partial}{\partial L^j} \left(\frac{(\vec{S_1}\cdot \vec{L})(\vec{S_2}\cdot \vec{L})}{L^5}\right)A^k \non\\
&=  \epsilon_{ijk} \left[ \frac{(\vec{S_2}\cdot \vec{L})}{L^5}S_{1j} + \frac{(\vec{S_1}\cdot \vec{L})}{L^5}S_{2j}-5(\vec{S_1}\cdot \vec{L}) (\vec{S_2}\cdot \vec{L})\frac{L_j A^k}{L^7} \right]~, \label{Ass2}
\end{align}
inserting $(\ref{Ass1})$ and $(\ref{Ass2})$ in $(\ref{Ass})$ we find
\be
\langle \dot{\vec{A}} \rangle_{SS} = \frac{ -3(-2E)^{\frac{3}{2}}}{2L^4}\left[ (\vec{S_2}\cdot \vec{\hat{L}})\vec{S_1}+(\vec{S_1}\cdot \vec{\hat{L}})\vec{S_2} +[ (\vec{S_2}\cdot\vec{S_1})-5 (\vec{S_1}\cdot \vec{\hat{L}})(\vec{S_2}\cdot \vec{\hat{L}})] \vec{\hat{L}} \right] \times \ \vec{A}~.
\label{SS-Adot}
\ee
Thus, the spin-spin coupling induce a rotation of the LRL vector $\vec{A}$, with an angular velocity
\be
\vec{\Omega}_{A_{SS}} = \frac{ -3(-2E)^{\frac{3}{2}}}{2L^4}\left[ (\vec{S_2}\cdot \vec{\hat{L}})\vec{S_1}+(\vec{S_1}\cdot \vec{\hat{L}})\vec{S_2} +[ (\vec{S_2}\cdot\vec{S_1})-5 (\vec{S_1}\cdot \vec{\hat{L}})(\vec{S_2}\cdot \vec{\hat{L}})] \vec{\hat{L}} \right]~. \label{OmegaAss}
\ee
Altogether, the anomalies given in $(\ref{omegalss}, \ref{omegaS1ss}, \ref{omegaS2ss}, \ref{OmegaAss})$ reproduce the corresponding expressions in \cite{Barker1970} in the test mass limit.
\subsection{Conserved quantities}
H the Hamiltonian is conserved by construction so E is a constant of motion.\\
From (\ref{Ldotss}, \ref{S1dotss}, \ref{S2dotss}, \ref{SS-Adot}) we see that
\begin{itemize}
\item  if $\vec{S_1}$, $\vec{S_2}$ and $\vec{L}$ are all parallel (or antiparallel) then they are constant of the motion,
\item for general orientation, $\vec{S_1}$, $\vec{S_2}$ and $\vec{L}$ are not conserved but $ \vec{J} =\vec{L} +\vec{S_1}+\vec{S_2}$  the total angular momentum of the system is conserved,
\item the magnitude of $\vec{L}$, $\vec{S_1}$ and $\vec{S_2}$ are always constant since the vectors are just rotating,
\item $\vec{A}$ the LRL vector is not conserved, whatever the general orientation of $\vec{S}$, but its magnitude is constant since the vector is just rotating.
\end{itemize}

\subsection*{Acknowledgments}

It is a pleasure to thank Gerhard Sch\"{a}fer for a discussion of the averaging method and for comments on a draft and Simon Caron-Huot for a discussion of the interpretation of the LRL vector and for some comments on the draft. 

This research was supported by the Israel Science Foundation grant no. 812/11 and and it is part of the Einstein Research Project ``Gravitation and High Energy Physics", which is funded by the Einstein Foundation Berlin.

\appendix
\section{Computation of $\delta_j r^ i$ and $\delta_j p^ i$}
\label{app:LRLtransformation}
In this appendix we explicitly compute the generators of phase space symmetry corresponding to the LRL vector.
We wish to simplify (\ref{deltajri}, \ref{deltajpi})
\begin{align}
\{ r^j p^2,r^i \} &= -\frac{\partial ( r^j p^2)}{\partial p^k} \frac{\partial r^i}{\partial r^k}=-r^j 2p^a\delta_{ak} \delta_{ik} = -2r^jp^i~,\\
\{ p^j (p^l\cdot r^l),r^i\} &= -\frac{\partial ( p^j (p^l\cdot r^l)) }{\partial p^k} \frac{\partial r^i}{\partial r^k} = -[(p^l\cdot r^l)\delta_{jk}+p^j r^l \delta_{lk}]\delta_{ik}\\
&= -[(p^l\cdot r^l)\delta_{ji}+p^j r^i]~,\\
\{ \alpha \frac{r^j}{r},r^i \}&=0~,
\end{align}
and
\begin{align}
\{ r^j p^2,p^i \} &=\frac{\partial ( r^j p^2)}{\partial r^k} \frac{\partial p^i}{\partial p^k}=p^2 \delta_{jk}\delta_{ik} =p^2 \delta_{ij}~,\\
\{ p^j (p^l\cdot r^l),p^i\} &= \frac{\partial ( p^j (p^l\cdot r^l)) }{\partial r^k} \delta_{ik} = p^j p^l \delta_{lk}\delta_{ik}= p^j p^i~, \\
\{\frac{r^j}{r},p^i \} &= \frac{\partial}{\partial r^k}(\frac{r^j}{r}) \delta_{ik} = [\frac {\delta_{jk}}{r}+r^j \frac{\partial}{\partial r^k}(r^m r^m)^{-\frac{1}{2}}] \delta_{ik}\\
&= (\frac {\delta_{ji}}{r}-\frac{r^j r^i}{r^3})~,
\end{align}
Finally substituting in (\ref{deltajri}, \ref{deltajpi}) we get
\begin{align} 
\delta_j r^ i &= \{A^j , r^i \}=[-2r^jp^i+r^i p^j +(\vec{r}\cdot\vec{p})\delta_{ij}]~,\\
\delta_j p^ i &= \{A^j , p^i \}=[p^2 \delta_{ij}-p^j p^i 
-\alpha (\frac {\delta_{ij}}{r}-\frac{r^j r^i}{r^3})]~. \label{DeltaLRL2}
\end{align}

\newpage
\section{Residue Method}
\label{app:Residu}
We want to calculate
\be
\int_{0}^{2\pi} d\theta \frac{1}{1+e\cos\theta}~, \label{residu1}
\ee
We change variables to the complex z plane:~$z=\exp{i\theta}$,~hence $dz=izd\theta$ and $\cos \theta =\frac{1}{2}(z+z^{-1})$ and substituting in ($\ref{residu1}$), the integration becomes a close contour complex integration $\oint f(z)dz$, with
\be
f(z)=\frac{1}{iz}\frac{1}{1+e(\frac{z+z^{-1}}{2})}=\frac{2}{ie} \frac{1}{z^2+\frac{2}{e}z+1}~,
\ee 
Cauchy theorem tells us that
\be
\oint f(z)dz=2\pi i \sum_{j=1}^{m}Res_{z_j}f ~,
\ee
with $z_j$ all the poles of $f(z)$ inside the integration contour
\begin{figure}[!hb]
\includegraphics[width=5cm]{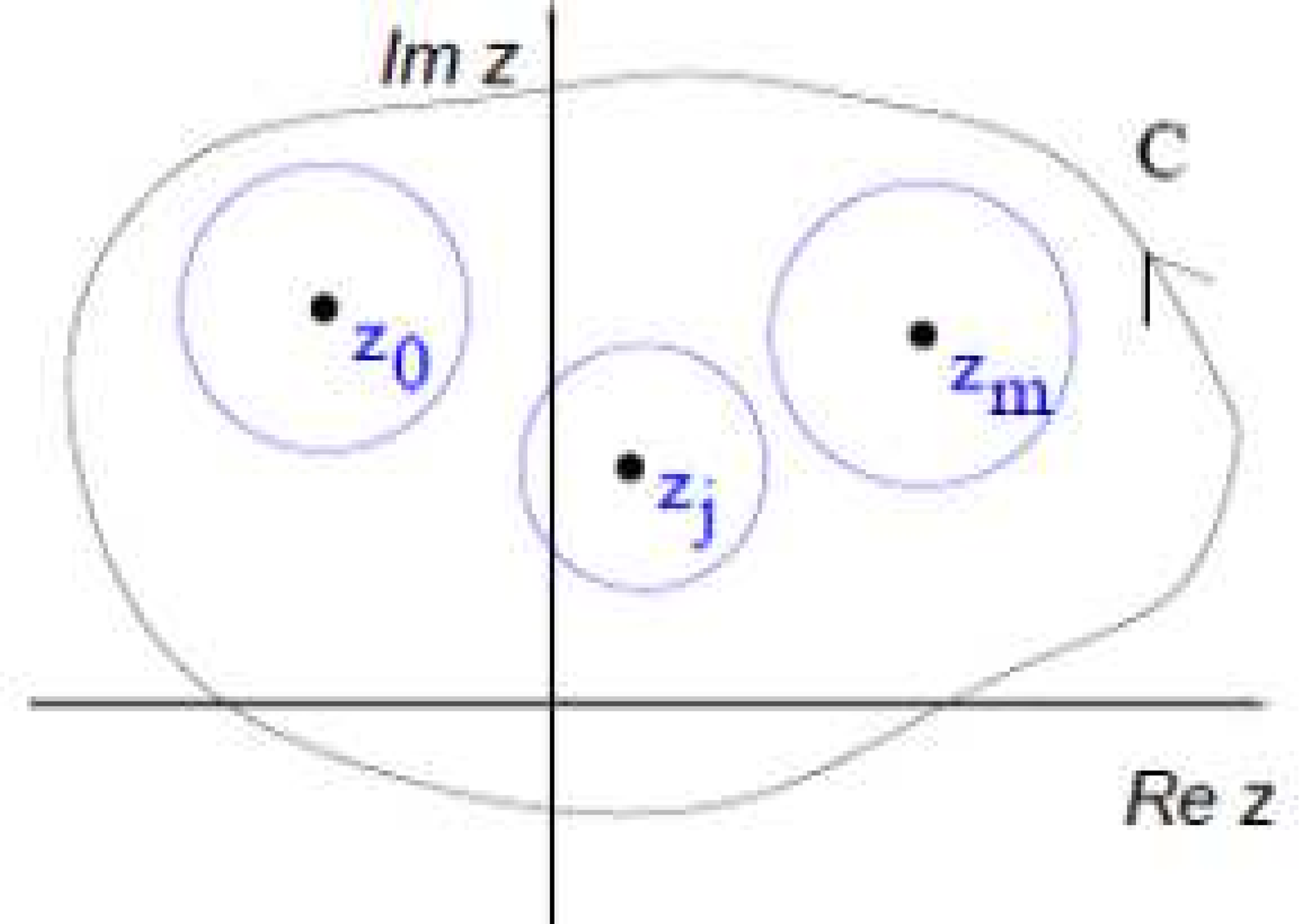}\centering
\end{figure}
\\
The poles of $f(z)$ are given by
\be
z_1=\frac{-1+\sqrt{(1-e^2)}}{e},~~~z_2=\frac{-1-\sqrt{(1-e^2)}}{e}~,
\ee
and
\be
f(z)=\frac{2}{ie}\frac{1}{(z-z_1)(z-z_2)}~,
\ee
\\
Since only $z_1$ is inside the integration contour ($e \leq 1$), we have to calculate
\be
Res_{z_1}f = \lim_{z \to z_1} ((z-z_1)f(z))=\frac{2}{ie}\frac{1}{z_1-z_2}=\frac{1}{i\sqrt{1-e^2}}~,
\ee
and finally
\be
\oint f(z)dz=2\pi i~ Res_{z_1}f = \frac{2\pi}{\sqrt{1-e^2}}~.
\ee

\end{document}